\documentclass[lettersize,journal]{IEEEtran}
\usepackage{amsmath, amsfonts, amsthm, bm}
\usepackage[T1]{fontenc}
\usepackage{algorithmic}
\usepackage{algorithm}
\usepackage{array}
\usepackage{textcomp}
\usepackage{url}
\usepackage{verbatim}
\usepackage{graphicx}
\usepackage{cite}
\usepackage{epsfig}
\usepackage{xcolor,soul}
\usepackage{multicol}
\usepackage{multirow}
\usepackage{amssymb}
\usepackage{hhline}
\usepackage{diagbox} 
\usepackage{tikz}
\usepackage{soul}

\begin{document}

\title{%
{\fontsize{22pt}{0pt}\selectfont
HQC Post-Quantum Cryptography Decryption with Generalized Minimum-Distance Reed-Solomon Decoder}%
}
\author{Jiaxuan Cai,~\IEEEmembership{Graduate Student Member,~IEEE} and Xinmiao Zhang,~\IEEEmembership{Fellow,~IEEE}
\thanks{
This material is based upon work supported by the National Science Foundation under Award No. 2052641. The authors are with the department of Electrical and Computer Engineering, The Ohio State University, Columbus, OH 43210 USA (e-mail: cai.1072@osu.edu; zhang.8952@osu.edu).}}

\markboth{ }%
{CAI AND ZHANG: Low-Complexity Parallel Min-Sum Medium-Density Parity-Check Decoder for McEliece Cryptosystem}
\maketitle

\begin{abstract}
Hamming Quasi-Cyclic (HQC) was chosen for the latest post-quantum cryptography standardization. A concatenated Reed-Muller (RM) and Reed–Solomon (RS) code is decoded during the HQC decryption. Soft-decision RS decoders achieve better error-correcting performance than hard-decision decoders and accordingly shorten the required codeword and key lengths. However, the only soft-decision decoder for HQC in prior works is an erasure-only decoder, which has limited coding gain. This paper analyzes other hardware-friendly soft-decision RS decoders and discovers that the generalized minimum-distance (GMD) decoder can better utilize the soft information available in HQC. Extending the Agrawal–Vardy bound for the scenario of HQC, it was found that the RS codeword length for HQC-128 can be reduced from 46 to 36. This paper also proposes efficient GMD decoder hardware architectures optimized for the short and low-rate RS codes used in HQC. The HQC-128 decryption utilizing the proposed GMD decoder achieves 20\% and 15\% reductions on the latency and area, respectively, compared to the decryption with hard-decision decoders.
\end{abstract}


\section{Introduction}
The National Institute of Standards and Technology (NIST) launched the post-quantum cryptography standardization \cite{NIST} in recent years. After evaluating the candidates in the latest round \cite{BIKENIST, NewBIKE, MDPCMcEliece, optLLMS, McElieceSlicedMP,HQC, LLMS}, the Hamming Quasi-Cyclic (HQC) was selected. The decryption of HQC first carries out a polynomial multiplication and a subtraction, and then performs concatenated Reed–Muller (RM) and Reed–Solomon (RS) decoding. The overall HQC system implementation has been explored in \cite{optHQC, HQCImpXu, HQCImpWang, HQCImpPelosi, HQCImpAlgoOpt, HQCImpPelosi2}. Optimizations have been carried out on data flow \cite{HQCImpXu} and memory \cite{optHQC, HQCImpAlgoOpt}. The complexity of polynomial multiplication is reduced by exploiting the sparsity of the operands in \cite{HQCImpWang, HQCImpPelosi}. A reconfigurable architecture supporting each HQC security level is presented in \cite{HQCImpPelosi2}.  Nevertheless, hard-decision RS decoding is utilized in all of these designs.  

Soft-decision decoding of RS codes can achieve lower decoding failure rate (DFR) than hard-decision decoding. Accordingly, it allows shorter codewords to be used to achieve the required security level and this means shorter key length. The only HQC decryption design with soft-decision RS decoding was proposed in \cite{HQCISCAS}. The intermediate results of RM decoding were used to find the least reliable symbols and an erasure-only RS decoder is employed. However, this decoding fails whenever the erased symbols do not include all the errors, and its achievable coding gain is limited. 

For the first time, this paper proposes a generalized minimum-distance (GMD) RS decoder design for HQC decryption. It carries out decoding on a number of error-and-erasure test vectors, and the erasure-only vector decoded in \cite{HQCISCAS} is one of them. Through analyzing the soft information available from RM decoding as well as simulations, it was discovered that the GMD decoding achieves the lowest DFR among soft-decision RS decoders with practical hardware complexity in the HQC setting. To derive the minimum codeword length needed to achieve the DFR required by the HQC, this paper also adapts the Agrawal-Vardy bound \cite{AV} for the concatenated RM and RS code structure. Last but not the least, efficient GMD decoder hardware implementation architectures are developed for the short and low-rate RS codes used in HQC. For HQC-128, the proposed design reduces the codeword length by 22\%.  The overall HQC-128 decryption utilizing the proposed GMD decoder achieves 20\% and 15\% reductions on the latency and area, respectively, compared to the decryption utilizing hard-decision decoders.

The organization of this paper is as follows. Section II introduces the background information. Section III presents the GMD decoder for HQC and Section IV derives the minimum required codeword length. Efficient GMD decoder hardware architectures and complexity comparisons are provided in Section V. Conclusions follow in Section VI.

\section{Background}
A ciphertext or secret key of the HQC public key encryption scheme consists of two polynomials in the quotient ring $\mathcal{R}=GF(2)[X]/(X^n-1)$. Denote the secret key by $\mathbf{sk}=(\mathbf{x}, \mathbf{y})$. The decryption of the ciphertext $(\mathbf u, \mathbf v)$ first computes $\mathbf c'=\mathbf{v} - \mathbf{u}\mathbf{y}$ and then carries out RM decoding followed by RS decoding on $\mathbf c'$. Randomly generated vectors have been multiplied and added to the RS and RM encoded message polynomial to derive the ciphertext during the encryption. If $\mathbf c'$ is corrupted by a correctable error vector, then the decoding yields the correct message polynomial.  
\begin{table}[!t]
    \renewcommand\arraystretch{1.1}
    
    \caption{HQC Parameters \cite{HQC}}\label{param}
    \vspace{-1.7em}
    \begin{center}
    \begin{tabular}{@{}c@{\hspace{1pt}}|@{}c@{}|@{}c@{}|@{}c@{}|@{}c@{}}
    \hline
    Security level $\lambda$& RM code & RM codeword copies $m$ & RS code &$n$  \\\hline\hline
    128 & $(128, 8)$ &3 & $(46, 16)$  & 17669\\\hline
    192 & $(128, 8)$ & 5& $(56, 24)$  & 35851 \\\hline
    256 & $(128, 8)$ & 5& $(90, 32)$ & 57637 \\\hline
    \end{tabular}
    \end{center}
    \vspace{-3em}
\end{table}

The length and dimension of a RM (RS) code are denoted by $(n_{RM(RS)}, k_{RM(RS)})$. The code parameters for HQC of security levels $\lambda$=128, 192, and 256 are listed in Table \ref{param}. During encryption, the RM codewords are repeated $m$ times. $n$ is set as the smallest prime larger than $mn_{RM}n_{RS}$ to avoid algebraic attacks \cite{HQC}. During decryption, the lowest $mn_{RM} n_{RS}$ bits of $\mathbf{c}'$ are first partitioned into $n_{RS}$ segments, each of length $mn_{RM}$-bit. The $mn_{RM}$ bits are further divided into $m$ groups. The `1' and `0' in each group are mapped to +1 and -1, respectively, and the values in the $m$ groups are added up component-wise to generate the inputs to the RM decoder. Each RM decoding outputs a $k_{RM}$-bit value, which is interpreted as one symbol over finite field $GF(2^{k_{RM}})$. After all $n_{RS}$ RM decoding is finished, the resulting $n_{RS}$ symbols form the input vector over $GF(2^{k_{RM}})$ for the RS decoder, whose output symbols are converted to binary format as the decrypted message.

For the RM code, efficient decoding can be achieved by a fast Hadamard transform (FHT)-based algorithm \cite{FHT}. It calculates a set of $2^{k_{RM}-1}=n_{RM}$ integer values, each representing the correlation between the received vector and a valid RM codeword. Then, the $k_{RM}$-bit decoding output is the concatenation of the sign bit and index of the entry with the largest magnitude. The conventional hard-decision RS decoding can correct $t=(n_{RS}-k_{RS})/2$ errors, and it is composed of three major steps: syndrome computation, key equation solver (KES) for obtaining the error locator and evaluator polynomials, and Chien search for identifying the error locations and computing the magnitudes. The syndrome computation and Chien search are implementable by constant finite field multipliers and adders \cite{HQCISCAS}. The most efficient KES architecture is the enhanced parallel inversionless Berlekamp-Massey (ePIBM) architecture \cite{Wu}. For an $(n_{RS}, k_{RS})$ RS code, a parallel ePIBM architecture consists of $2t+1$ identical processing elements and has a latency of $2t$ clock cycles.


For HQC of security level $\lambda$, the product of the DFRs of RM and RS decoding must not exceed $2^{-\lambda}$ to avoid attacks that exploit decoding failures \cite{HQC}. $k_{RS}$ is set to $\lambda/k_{RM}$. To reduce the key size of HQC, $2n$, the RS code with the smallest $n_{RS}$ achieving the DFR requirement should be utilized. The parameters listed in Table \ref{param} are decided based on hard-decision RS decoding. On the other hand, the magnitudes of the integers computed by the FHT serve as metrics of reliability. They can be utilized to carry out soft-decision RS decoding, which achieves better error-correcting performance and hence needs smaller $n_{RS}$ to reach the same DFR as hard-decision decoding. 

\section{GMD RS Decoding for HQC}
This section investigates hardware-friendly soft-decision RS decoding algorithms, including erasure-only, Chase, and GMD decoding, for HQC decryption to reduce the RS codeword length. Different from those in conventional communication systems, the soft inputs to the RS decoder are generated from RM decoding in HQC. From analyses and confirmed by simulations, it will be shown that the GMD decoding achieves the lowest DFR in this scenario.
\begin{figure}[t] 
    \begin{center}
    \includegraphics[width=.43\textwidth]{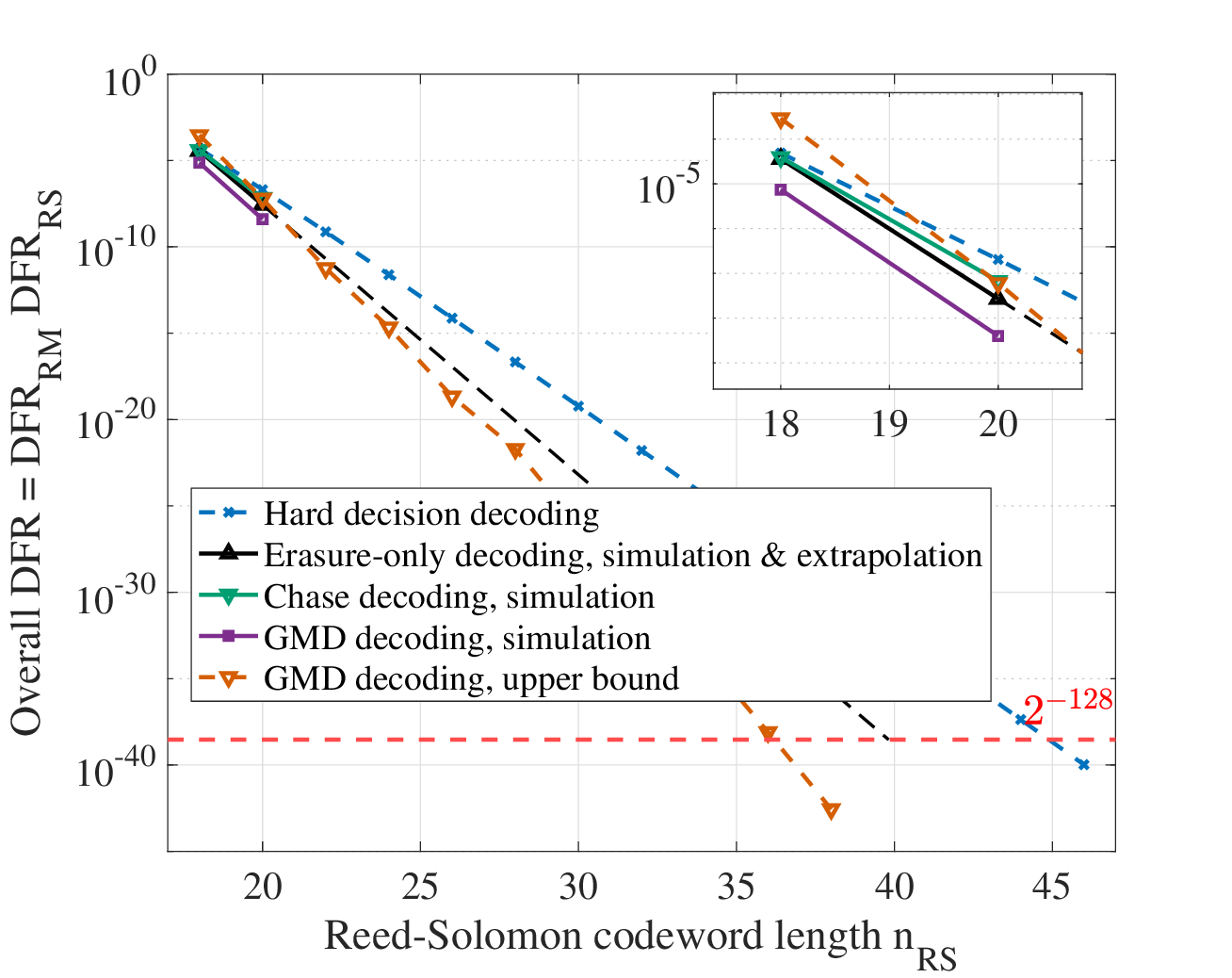}
    \vspace{-.5em}
    \caption{DFRs of the concatenated RM $(128, 8)$ and RS $(n_{RS}, 16)$ code for HQC-128.}\label{extrapolation}
    \end{center}
    \vspace{-2em}
\end{figure}

The FHT for RM decoding computes a set of $2^{k_{RM}-1}=n_{RM}$ integers. As mentioned previously, the sign and index of the integer with the largest magnitude form an input symbol to the RS decoder. This largest magnitude, denoted by $max_1$, can be utilized as the reliability of the symbol. Erasure-only decoding is exploited in \cite{HQCISCAS}. An $(n_{RS},k_{RS})$ RS code can correct up to $n_{RS}-k_{RS}=2t$ erasures and the design in \cite{HQCISCAS} sets the $2t$ symbols with the smallest $max_1$ as erasures. It succeeds only when the erasure set covers all the errors. Hence, its achievable coding gain  over hard-decision decoding is limited as shown by the simulation and extrapolation results in Fig. \ref{extrapolation}. For HQC with $\lambda=128$, it reduces the required RS codeword length from 46 to 40.

The Chase decoding \cite{Chase} tries the two most likely symbols for each of the $\eta$ least reliable code positions. In total, it carries out $2^{\eta}$ decoding trials. To reduce the complexity, a small $\eta$ is typically used. Similar to that in erasure-only decoding, the $\eta$ least reliable positions can be decided based on the $max_1$ values. From RM decoding, the best possible guess of the second most likely symbol is the one corresponding to the integer with the second largest magnitude among all the $n_{RM}$ integers computed, denoted as $max_2$. However, from simulations over $10^9$ RM decoding samples, it was found that the correct symbol does not correspond to either $max_1$ or $max_2$ with a probability of $2.39\times 10^{-4}$. The Chase decoding may correct more errors than hard-decision decoding only when the second most likely symbols in some of the $\eta$ least reliable positions are actual correct symbols. As a result, the Chase decoding with $\eta=3$ has even higher DFR than erasure-only decoding as shown in Fig. \ref{extrapolation}.

To further reduce the DFR, this paper proposes to use the GMD decoding \cite{GMD}, which carries out $t+1$ decoding trials. In   the $i$-th trial ($0\leq i \leq t$), the $2i$ symbols with the smallest $max_1$ values are assigned as erasures. The erasure-only decoding is the last decoding trial in the GMD decoding. The decoding trial with $2i$ erasures can correct $t-i$ extra errors outside of the erasure set. If some errors fail to be included in the erasure set, they can still potentially be corrected in such error-and-erasure decoding. As shown in the simulation results in Fig. \ref{extrapolation}, the GMD decoding achieves noticeable DFR reduction compared to the Chase and erasure-only decoding for HQC even for the short $(18,16)$ and $(20,16)$ RS codes.

\section{Upper Bound of GMD DFR for HQC}
Simulations for reaching the target DFR of $2^{-\lambda}$ required by HQC cannot be completed in reasonable time. This section derives an upper bound of the DFR of the proposed GMD decoding for HQC in order to determine the minimum RS codeword length that achieves the target DFR. The derivation incorporates the concatenated RM and RS structure of HQC into the probabilistic analyses of the Agrawal–Vardy bound \cite{AV}, which is tighter than the bound from \cite {GMD}.



Let $\Pr(\mathcal{F}_i)$ denote the probability that the $i$-th decoding trial with $2i$ erasures in GMD fails ($0 \leq i\leq t$). The overall GMD decoding fails only if all trials fail, and its probability is  $\Pr(\mathcal{F}_{GMD})=\prod_{0 \leq i\leq t} \Pr(\mathcal{F}_i)$. Since each $\Pr(\mathcal{F}_i)<1$, $\Pr(\mathcal{F}_{GMD})<\min_{i} \Pr(\mathcal{F}_i)$. Let $E_i$ be the number of errors outside of the erasure set in the $i$-th decoding trial. Then $\Pr(\mathcal{F}_i)=\Pr(E_i>t-i)$. In \cite{AV}, it was assumed that the soft inputs to the GMD decoder are independent and identically distributed (i.i.d.). Hence, $E_i$ follows a binomial distribution, which is approximated as $B(n_{RS} q_i, {\epsilon}_i)$. Here, $q_i$ denotes the probability that a decoder input symbol is outside of the erasure set and $\epsilon_i$ is the probability that a symbol outside of the erasure set is erroneous in the $i$-th decoding trial. Formulas for computing $q_i$ and $\epsilon_i$ are given in \cite{AV} for the case that the RS decoder inputs come from AWGN channel. Since $E_i$ follows binomial distribution, the Chernoff bound in Kullback-Leibler divergence form \cite{KL} can be applied to derive
\begin{equation}
\Pr(\mathcal{F}_i) \le \exp\left(-n_{RS} q_i D\left(\frac{t-i}{n_{RS} q_i}\Big\Vert {\epsilon}_i\right)\right),
\label{eq:single_trial_fail_bound}
\end{equation}
where $D(x\Vert y)=x\ln\frac{x}{y}+(1-x)\ln\frac{1-x}{1-y}$ is the Kullback--Leibler divergence. An upper bound of the GMD decoding failure is the minimum of the values in \eqref{eq:single_trial_fail_bound} over $0\leq i\leq t$.

In the HQC scenario, the soft inputs of RS decoder come from RM decoding. Although the inputs are discrete instead of continuous values as in the case of AWGN channel, they are also i.i.d. Hence, the upper bound in \eqref{eq:single_trial_fail_bound} still applies. However, the formulas for $q_i$ and $\epsilon_i$ presented in \cite{AV} for AWGN can not be used. It is difficult to develop closed-form formulas for $q_i$ and $\epsilon_i$ from RM decoding results. Instead, these values can be derived from simulations over a large number of samples, such as $10^9$. Plugging them into \eqref{eq:single_trial_fail_bound}, the upper bounds of the DFRs for GMD decoding of different codeword lengths can be derived. The DFR upper bounds of HQC-128 with RS codes with different $n_{RS}$ but the same $k_{RS}=16$ are plotted in Fig. \ref{extrapolation}. It can be observed that $n_{RS}=36$ is sufficient to reach a DFR below $2^{-128}$ when GMD decoding is employed. As a result, $n$ is reduced from 17669 as listed in Table \ref{param} to $13829$, which is the smallest prime larger than $mn_{RM}  n_{RS}=13824$. This leads to a key length reduction of 1-13829/17669=22\%.

\section{Hardware Architectures and Comparisons}
This section proposes efficient hardware architectures for GMD RS decoding optimized for the short and low-rate codes used in HQC. The hardware complexity is analyzed and compared with those of prior designs using erasure-only decoding \cite{HQCISCAS} and hard-decision decoding \cite{HQCImpWang,HQCImpPelosi}.

In HQC decryption, $\mathbf{c}'=\mathbf{v} - \mathbf{u}\mathbf{y}$ is first computed. The polynomial multiplication can be implemented using the shift-and-add architectures proposed in \cite{HQCImpWang, HQCImpPelosi}. Then, the lowest $mn_{RM}\cdot n_{RS}$ coefficients of $\mathbf{c}'$ are split into $n_{RS}$ segments of $m$ groups, each of which has $n_{RM}$ bits. These bits are mapped to +1 or -1  and the values from the $m$ groups are added up component wise to generate  $n_{RM}$ integers, which form the inputs to the RM decoder. In the RM decoding, the inputs first go through FHT, which computes a vector of $n_{RM}$ integers. The FHT can be implemented using the iterative butterfly architecture in \cite{RMImpl}, which has $\log_{2} n_{RM}$ stages of $n_{RM}$ adders. Then a binary tree finds the $max_1$ value in the FHT output vector and its index, which is concatenated with the sign bit of $max_1$ to form the RM decoder output symbol. One RM decoder is employed in our design. It carries out $n_{RS}$ RM decoding one after another. The $max_1$ value of each RM decoding is sent to a sorter that orders the $2t$ smallest values. The sorter can be implemented using the  insertion machine in \cite{sorter} with $2t$ cells, each consisting of a register, a multiplexer, and a comparator. Since each RM decoding takes multiple clock cycles, the folding technique can be applied to reduce the sorter area.

\begin{figure}[t] 
    \begin{center}
    \includegraphics[width=.48\textwidth]{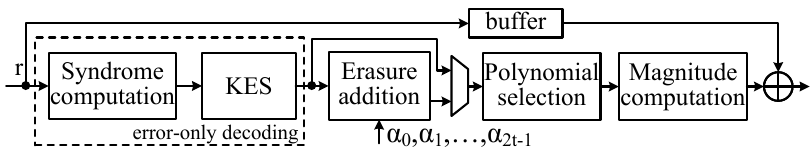}
    \vspace {-0.5em}\caption{Block diagram of the one-pass GMD RS decoder.}
\label{overall_GMD}
    \end{center}
    \vspace{-2.5em}
\end{figure}

The vector of $n_{RS}$ RM output symbols, denoted by $r=[r_{n_{RS}-1}, \cdots, r_1, r_0]$, along with the ordered list of the $2t$ least reliable symbols, denoted by $\alpha_0,\alpha_1,\cdots,\alpha_{2t-1}$, are the inputs to the GMD decoder, whose block diagram is shown in Fig. \ref{overall_GMD}. First, error-only decoding is carried out on $r$. Define $r(X)=r_{n_{RS}-1}X^{r_{RS}-1}+\cdots+r_1X+ r_0$. The syndrome computation is to compute $S_j=r(\alpha^j)$ ($1\leq j\leq 2t$), where $\alpha$ is a primitive element of $GF(2^{k_{RM}})$. Applying the Horner's rule, each syndrome can be computed by a feedback loop consisting of a constant multiplier, an adder, and a register. $2t$ units of such feedback loops compute all the syndromes in $n_{RS}$ clock cycles. Then the ePIBM architecture \cite{Wu} can be utilized to compute the error locator polynomial $\Lambda(X)$ and its auxiliary polynomial $B(X)$. The fully parallel ePIBM architecture consists of $2t+1$ processing elements, each of which has two multipliers, one adder, two multiplexers, and two registers. It can complete the KES step in $2t$ clock cycles. Similarly, the folding technique can be adopted to reduce the area at the cost of increasing the latency. The polynomial selection block in Fig. \ref{overall_GMD} carries out Chien search on $\Lambda(X)$, which is implemented by an array of constant multipliers \cite{ZhangGMD}. If its root number equals its degree, then the error-only decoding is considered successful. If $\Lambda(\alpha^{-l})=0$, then $r_{l}$ is erroneous, and the error magnitude is computed by using the Horiguchi–K\"otter formula \cite{Wu} as 
\vspace{-0.5em}
\begin{equation}
e_l=\gamma\Lambda_0\alpha^{-l\cdot2t}/B(\alpha^{-l})\Lambda_{odd}(\alpha^{-l}),
\label{eq:errmag}
\vspace{-0.5em}
\end{equation}
where $\gamma$ is a value from the last iteration of the ePIBM algorithm and $\Lambda_0$ is the constant coefficient of $\Lambda(X)$. $B(\alpha^{-l})$ can also be computed by the Chien search architecture.

\begin{algorithm}[!t]
\caption{One-Pass GMD Erasure Addition Algorithm \cite{WuGMD}}
\label{WuGMDalgo}
\begingroup
\setlength{\baselineskip}{0.8\baselineskip}
\begin{algorithmic}[1]
\small
\STATE \textbf{Input:}  $\Lambda(X),B(X)$, $(\alpha_0,\alpha_1,\ldots,\alpha_{2t-1})$.
\STATE \textbf{Initialization:} $\Lambda^{(0)}(X) \gets \Lambda(X)$,\quad $\mathcal{B}^{(0)}(X) \gets XB(X)$.
\STATE \textbf{for} $i=0,\ldots,2t-1$ \textbf{do}
    \STATE $\quad$ \textbf{Polynomial evaluation:}
    \STATE $\quad$ $\Lambda_{i} \gets \Lambda^{({i})}(\alpha_i^{-1})$,\quad $\mathcal{B}_{i} \gets \mathcal{B}^{({i})}(\alpha_i^{-1})$
    \STATE $\quad$ \textbf{Polynomial updating:}
    \STATE $\quad$ $\quad$ \textbf{if} $\Lambda_{i}=0 \ \vee\ \big(\Lambda_{i}\neq 0 \wedge \mathcal{B}_{i}\neq 0 \wedge L_{\Lambda}^{(i)} \ge L_{\mathcal{B}}^{(i)}\big)$:

    \STATE $\quad$ $\quad$$\quad$$\Lambda^{(i+1)}(X) \gets \mathcal{B}_{i}\cdot \Lambda^{({i})}(X) - \Lambda_{i} \cdot \mathcal{B}^{({i})}(X)$
    \STATE $\quad$ $\quad$$\quad$$\mathcal{B}^{(i+1)}(X) \gets (X-\alpha_i^{-1})\,\mathcal{B}^{({i})}(X)$
    \STATE $\quad$ $\quad$$\quad$$L_{\Lambda}^{(i+1)} \gets L_{\Lambda}^{(i)}, L_{\mathcal{B}}^{(i+1)} \gets L_{\mathcal{B}}^{(i)}+1$

    \STATE $\quad$ $\quad$  \textbf{else if} $\mathcal{B}_{i}=0 \ \vee\ (\Lambda_{i}\neq 0 \wedge \mathcal{B}_{i}\neq 0 \wedge L_{\Lambda}^{(i)}<L_{\mathcal{B}}^{(i)})$:

    \STATE $\quad$ $\quad$$\quad$$\Lambda^{(i+1)}(X) \gets (X-\alpha_i^{-1})\,\Lambda^{({i})}(X)$
    \STATE $\quad$ $\quad$$\quad$$\mathcal{B}^{(i+1)}(X) \gets \mathcal{B}_{i}\cdot X\Lambda^{({i})}(X) - \alpha_i^{-1}\Lambda_{i} \cdot \mathcal{B}^{({i})}(X)$
    \STATE $\quad$ $\quad$$\quad$$L_{\Lambda}^{(i+1)} \gets L_{\Lambda}^{(i)}+1, L_{\mathcal{B}}^{(i+1)} \gets L_{\mathcal{B}}^{(i)}$

\end{algorithmic}
\endgroup
\end{algorithm}

Instead of carrying out the KES for each error-and-erasure vector separately, the algorithms in \cite{GMDKES,WuGMD,ZhangGMD} can be adopted to derive the locator and auxiliary polynomials of each test vector in one run. The one-pass GMD scheme in \cite{ZhangGMD} employs the interpolation-based RS decoding. Although it achieves significant savings over the K\"{o}tter's one-pass GMD \cite{GMDKES} for high-rate RS codes, it would involve three instead of two bivariate polynomials and require an extra complicated factorization step for the low-rate codes utilized in HQC. Both the K\"otter's \cite{GMDKES} and Wu's \cite{WuGMD} schemes start from the error-only decoding and the ePIBM architecture can be utilized. Compared to K\"otter's design, Wu's one-pass GMD procedure has the advantage that it can start directly from the ePIBM results instead of requiring another polynomial multiplication step. This algorithm is listed in Algorithm \ref{WuGMDalgo}. Here, $\Lambda^{(i+1)}(X)$ and $\mathcal{B}^{(i+1)}(X)$ denote the updated polynomials in the $i$-th iteration, and they are initialized as $\Lambda^{(0)}(X)=\Lambda(X)$ and $\mathcal{B}^{(0)}(X)=XB(X)$, respectively. $L_{\Lambda}^{(i)}$ and $L_{\mathcal{B}}^{(i)}$ denote the lengths of $\Lambda^{(i)}(X)$ and $\mathcal{B}^{(i)}(X)$, respectively.

\begin{figure}[t]
\vspace{-1.2em}
\centering
\includegraphics[width=.87\columnwidth]{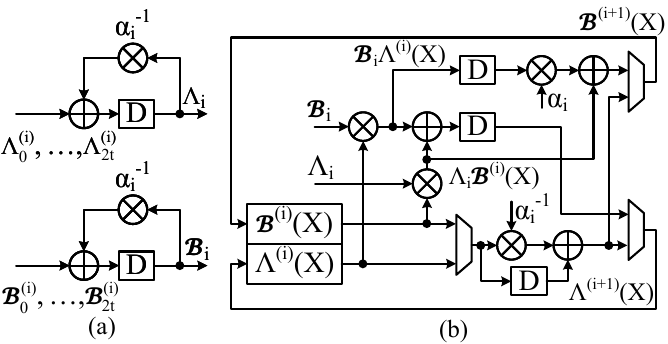}
\vspace{-.5em}
\caption{Implementation architectures for GMD erasure addition: (a) polynomial evaluation, (b) polynomial updating.}
\label{onepassimpl}
\vspace{-2em}
\end{figure}

\begin{figure}[t]
\vspace{-1.2em}
\centering
\includegraphics[width=1\columnwidth]{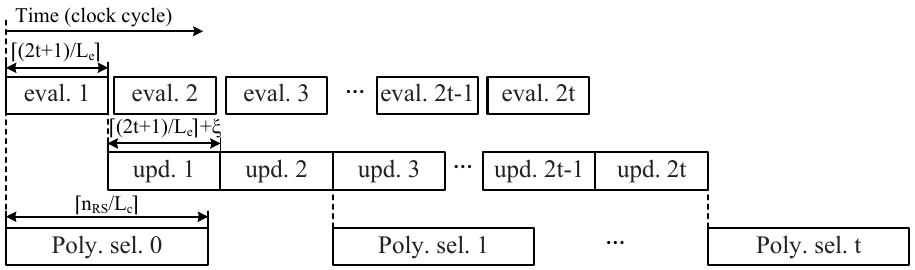}
\caption{Computation scheduling of the proposed erasure addition and polynomial selection.}
\label{GMDtiming}
\vspace{-2em}
\end{figure}

For the first time, efficient implementation architectures are proposed in this paper for Algorithm \ref{WuGMDalgo} as shown in Fig. \ref{onepassimpl}. First, $\Lambda_i$ and $\mathcal{B}_{i}$ can be computed by using two multiplier-adder feedback loops as shown in Fig. \ref{onepassimpl}(a) according to the Horner's rule in $2t+1$ clock cycles. The polynomial updating in rows 8-9 and 12-13 of Algorithm \ref{WuGMDalgo} can be implemented by the serial architecture in Fig. \ref{onepassimpl}(b) starting from the most significant coefficients. Line 9 and 12 can share the multiplier, adder, and register on the bottom of Fig. \ref{onepassimpl}(b) to implement the multiplication with $(X-\alpha_i^{-1})$. The adder and two multipliers in the left part of Fig. \ref{onepassimpl}(b) compute ${\Lambda}^{(i+1)}(X)$ in Line 8. The intermediate results, $\mathcal{B}_i{\Lambda}^{(i)}(X)$ and $\Lambda_i\mathcal{B}^{(i)}(X)$, can be re-used for computing $\mathcal{B}^{(i+1)}(X)$ in Line 13. The multiplication of $\alpha_i^{-1}$ with $\Lambda_i\mathcal{B}^{(i)}(X)$ in this line would lead to one extra multiplier in the critical path. To solve this issue, this line can be scaled by $\alpha_i$ to become $\mathcal{B}^{(i+1)}(X)\gets\alpha_i\cdot \mathcal{B}_i\cdot X\Lambda^{({i})}(X) - \Lambda_i \cdot \mathcal{B}^{({i})}(X)$. The multiplication of $\mathcal{B}_i\Lambda^{({i})}(X)$ by $X$ means that a delay element is added after $\mathcal{B}_i\Lambda^{({i})}(X)$. This delay separates the multiplication of $\alpha_i$ from the data path. Because of this scaling, $\mathcal{B}^{(i+1)}(X)$ and $\Lambda^{(i+1)}(X)$ may have different scaling factors. This can be compensated by modifying the Horiguchi-K\"otter formula for error/erasure magnitude computation.

The scheduling of the computations in the erasure addition and polynomial selection is shown in Fig. \ref{GMDtiming}. The degree of $B (X)$ from the ePIBM algorithm for error-only decoding can be as high as $2t$. Hence, each iteration of Algorithm 1 needs to process polynomails with degree up to $2t$. In an $L_e$-parallel design, the polynomial evaluation for each iteration takes $\lceil (2t+1)/L_e\rceil$ clock cycles. The polynomial updating units have $\xi=2$ pipelining stages. Hence, it takes $\lceil (2t+1)/L_e\rceil+\xi$ clock cycles to finish in an $L_e$-parallel design. However, both the polynomial evaluation and updating process the coefficients in the order of the most significant to the least significant. Hence, the polynomial evaluation for the next iteration of Algorithm 1 can start after the first pair of coefficients are updated in the current iteration.  

In Fig. \ref{GMDtiming}, the polynomial selection for vector 0 is for the error-only decoding. For the one-pass GMD process, polynomial selection is carried out on $\Lambda(X)$ to count its root number once after every two erasure addition iterations. To match the speed of the erasure addition, the Chien search needs to be completed in $2(\lceil (2t+1)/L_e\rceil+\xi)$ clock cycles and its parallelism, $L_c$, can be set accordingly. The $\Lambda(X)$ whose root number equals its degree is considered to be the correct error-and-erasure locator polynomial. The coefficients of $\Lambda(X)$ are split into the even and odd parts. The $n_{RS}$ outputs of the Chien search on the odd parts are stored to be utilized in the error/erasure magnitude computation using the Horiguchi-K\"otter formula. The $ B(\alpha^{-l})$ in this formula can be computed by reusing the Chien search engine.

The parallel processing/folding factors of different blocks can be adjusted to trade off area and latency. Finite field multipliers are expensive hardware units. Each multiplier over $GF(2^8)$ needs the area of 100 XOR gates to implement \cite{ZhangGMD}. The parallel ePIBM architecture requires $2(2t+1)$ multipliers and has a short latency of $2t$ clock cycles. On the other hand, the serial erasure addition architecture in Fig. \ref{onepassimpl} has long latency and requires a small number of multipliers. The overall HQC decryption latency can be effectively reduced by applying folding to the ePIBM architecture and using the saved multipliers to achieve parallel implementation of erasure addition. For HQC-128, it is found that using a folding factor of 3 on the ePIBM architecture and parallel-processing factor of $L_e=6$ on the erasure addition leads to the best tradeoff. From $L_e=6$, choosing $L_c=3$ for the Chien search avoids data buildup. Table \ref{tab:rs_decoder_breakdown} lists the hardware requirement and latency of each
module in the proposed GMD RS decoder.

\begin{table}[!t]
\centering
\renewcommand\arraystretch{1 }
\caption{Hardware Requirement and Latency of the Proposed GMD Decoder for RS(36, 16) Code over $GF(2^8)$}
\vspace{-1em}
\label{tab:rs_decoder_breakdown}
\begin{tabular}{@{}c@{\hspace{1pt}}|@{\hspace{1pt}}c@{\hspace{1pt}}|@{\hspace{1pt}}c@{\hspace{1pt}}|@{\hspace{1pt}}c@{\hspace{1pt}}|@{\hspace{1pt}}c@{\hspace{1pt}}|@{\hspace{1pt}}c@{\hspace{1pt}}|@{\hspace{1pt}}c@{\hspace{1pt}}||@{\hspace{1pt}}c@{\hspace{1pt}}}
\hline
 & General & Adder & Inverter & Constant & Mux & Register  & Latency \\
  & multiplier &  &  & multiplier & (8-bit) & (8-bit) & (\# of clks) \\
\hline
Syn. comp.        & 0  & 20 & 0 & 20 & 0   & 20  & 36  \\\hline
KES & 14 & 7 & 0 & 0  & 28 & 56  & 60  \\\hline
Erasure add. & 36  & 30 & 0 & 0 &  18  & 20  & 124  \\\hline
Poly. sel. & 0  & 63 & 0 & 60 & 20   & 56  & 12  \\\hline
Mag. comp.& 5  & 0 & 1 & 0 & 21   & 1  & 36  \\\hline
\hline
Total  &  55 & 120 & 1 & 80& 87 & 153 &  268  \\
\hline
\end{tabular}
\vspace{-.5em}
\end{table}


\begin{table}[t]
\centering
\renewcommand\arraystretch{1}
\caption{Area and Latency Comparison for HQC-128 Decryption using GlobalFoundries 22FDX Process with $320\ ps$ Timing Constraint}
\vspace{-1em}
\label{tab:overall_comparison}
\begin{tabular}{@{\hspace{1pt}}c@{\hspace{1pt}}| c | c | c }
\hline
& Proposed GMD &   Erasure-only &  Hard-decision  \\
&  & \cite{HQCISCAS} &  \cite{HQCImpWang, HQCImpPelosi}  \\\hline \hline
RS code used &$(36,16)$  &  $(40,16)$ & $(46,16)$\\\hline
Key len. (bits)  & 13829  & 15361 & 17669\\
(normalized) &\textbf{(0.78)}  & \textbf{(0.87)} & \textbf{(1)}\\\hline\hline
Total memory (bits) & 28828 & 31900 & 36574 \\\hline
Logic area ($\mu m^2$) & 6908 & 6412 & 6523 \\\hline
Total area (\# of XOR)&39190 &41518&46359\\
(normalized)& \textbf{(0.85)} & \textbf{(0.90)} & \textbf{(1)}\\\hline\hline
Latency (\# of clks) & 8265& 9005 & 10309\\
(normalized)&\textbf{(0.80)}&\textbf{(0.87)}&\textbf{(1)}\\\hline
\end{tabular}
\vspace{-2em}
\end{table}

The HQC-128 decryption with the proposed GMD decoder, prior erasure-only decoder \cite{HQCISCAS}, and prior hard-decision decoder \cite{HQCImpWang, HQCImpPelosi} are synthesized using GlobalFoundries 22FDX process under the timing constraint of $320\ ps$, which is the tightest timing constraint without increasing the area substantially. The results are listed in Table \ref{tab:overall_comparison}. The polynomial operations adopt the designs from \cite{HQCImpWang, HQCImpPelosi} and their memory requirements are also included in the table. In the 22FDX process, one XOR gate has an area of 0.67 $\mu m^2$. Assuming one memory bit occupies the same area as one XOR gate \cite{ZhangGMD}, the total areas of the three designs in terms of equivalent XOR gates are listed in Table~\ref{tab:overall_comparison}. It can be observed that the HQC decryption with the proposed GMD decoder has a much smaller memory size and overall area compared to prior designs because of the shorter codeword length, despite that the GMD decoder is more complicated. For the overall HQC-128 decryption, our new design reduces the area and latency by 15\% and 20\%, respectively, compared to the designs in \cite{HQCImpWang, HQCImpPelosi} with hard-decision RS decoder.

For HQC-192 and 256, the codewords are duplicated by 5 instead of 3 times to form ciphertexts. Hence, a reduction on RS codeword length leads to more significant reduction on the ciphertext length. Since the polynomial operations dominate the overall latency and area of the decryption and they reduce almost linearly with shorter ciphertext length, our proposed scheme will lead to more substantial savings in the cases of HQC-192 and 256. 


\section{Conclusions}
This paper proposes to adopt the GMD decoder for HQC decryption to better utilize the available reliability information. The soft information generated by the RM decoding is incorporated to derive an upper bound of the GMD decoder performance at low DFR region. It was shown that the codeword length needed for HQC can be substantially reduced. Efficient hardware architectures are proposed for the GMD decoder for the short and low-rate RS codes in HQC. Our proposed design for decryption achieves significant latency and area reduction.

\end{document}